\documentclass[pra,twocolumn,showpacs,preprintnumbers,amsmath,amssymb,superscriptaddress]{revtex4-2}
\usepackage{graphicx}
\usepackage{dcolumn}
\usepackage{bm}
\usepackage{epsfig}
\usepackage{amsmath}
\usepackage{amsfonts}
\usepackage{amssymb}
\usepackage{empheq}
\usepackage{color}
\usepackage{bbm}
\usepackage[caption=false]{subfig}
\usepackage{placeins}
\usepackage{soul}
\usepackage{physics}
\usepackage{mathrsfs}
\usepackage{changes} 
\definechangesauthor[name={Per cusse}, color=red]{per} 
\usepackage[colorlinks=true,citecolor=blue,linkcolor=blue,urlcolor=blue]{hyperref}
\usepackage{orcidlink}
\usepackage{appendix}

\begin{document}
\title{Manipulating spectral transitions and photonic transmission in a non-Hermitian optical system through nanoparticle perturbations}
\author{Bo-Wang Zhang~\orcidlink{0000-0003-1223-3613}\textsuperscript{$\circ$}}
\affiliation{\mbox{Center for Quantum Sciences and School of Physics, Northeast Normal University, Changchun 130024, China}}
\author{Cheng Shang~\orcidlink{0000-0001-8393-2329}\textsuperscript{$\circ$}}
\email{\textcolor{black}{Corresponding author:} cheng.shang@riken.jp}
\affiliation{\mbox{Department of Physics, The University of Tokyo, 5-1-5 Kashiwanoha, Kashiwa, Chiba 277-8574, Japan}}
\affiliation{Analytical quantum complexity RIKEN Hakubi Research Team, RIKEN Center for Quantum Computing (RQC), 2-1 Hirosawa, Wako, Saitama 351-0198, Japan}
\author{J. Y. Sun}
\affiliation{\mbox{Center for Quantum Sciences and School of Physics, Northeast Normal University, Changchun 130024, China}}
\author{Zhuo-Cheng Gu~\orcidlink{0009-0007-1629-0750}}
\email{\textcolor{black}{Corresponding author:} gu@stat.phys.titech.ac.jp}
\affiliation{\mbox{Department of Physics, Institute of Science Tokyo, Meguro, Tokyo, 152-8551, Japan}}
\author{X. X. Yi~\orcidlink{0000-0002-7031-2988}}
\email{\textcolor{black}{Corresponding author:} yixx@nenu.edu.cn \\ \textcolor{black}{$ \circ $ These two authors contributed equally to this work.}}
\affiliation{\mbox{Center for Quantum Sciences and School of Physics, Northeast Normal University, Changchun 130024, China}}
\affiliation{Center for Advanced Optoelectronic Functional Materials Research \\and Key Laboratory for UV Light-Emitting Materials and Technology of Ministry of Education, Northeast Normal University, Changchun 130024, China}

\date{\today}
\begin{abstract}
  In recent years, extensive research has been dedicated to the study of parity-time ($\mathcal{PT}$) symmetry, which involves the engineered balance of gain and loss in non-Hermitian optics. Complementary to $\mathcal{PT}$ symmetry, the concept of anti-$\mathcal{PT}$ symmetry has emerged as a natural framework for describing the dynamics of open systems with dissipations. In this work, we study spectral transitions and photon transmission in a linear spinning resonator perturbed by nanoparticles. First, we show that by precisely controlling the nanoparticle perturbations, the eigenvalues (or spectra) of a non-Hermitian system satisfying anti-$\mathcal{PT}$ symmetry can transit to that of a quasi-closed Hermitian system. Second, we outline the essential conditions for constructing a quasi-closed system and analyze its dynamic behavior with respect to photon transmission. By adjusting the rotational angular velocity of the spinning resonator and the strength of the nanoparticle perturbations, the quasi-closed system enables a variety of photon distribution behaviors, which may have significant applications in quantum devices. Our findings offer valuable insights for the design of dissipative quantum devices under realistic conditions and for understanding their responses to external perturbations.
\end{abstract}

\maketitle

\section{Introduction}
Non-hermiticity has emerged as a powerful framework for describing open quantum systems, revealing exotic phase transitions and dynamic behaviors around exceptional points (EPs)~\cite{miri2019exceptional,PhysRevLett.103.093902,kato2013perturbation}. Notably, Carl M. Bender and Stefan Boettcher introduced the concept of parity-time ($\mathcal{PT}$) symmetry, characterized by a balance of gain and loss, within the context of optics, which has since attracted significant attention across the natural sciences~\cite{Bender1998,Bender_2007,el2018non}. In the $\mathcal{PT}$-symmetric phase, non-Hermitian Hamiltonians can exhibit entirely real spectra, a phenomenon that has been experimentally observed in various physical platforms, including optical microcavities~\cite{chang2014parity,peng2014parity}, atomic systems~\cite{PhysRevLett.117.123601,li2019observation}, and acoustic devices~\cite{PhysRevX.4.031042,fleury2015invisible}. When the system parameters surpass the EP, $\mathcal{PT}$ symmetry is spontaneously broken, leading to a phase transition from the $\mathcal{PT}$-symmetric phase to the $\mathcal{PT}$-symmetry-broken phase~\cite{PhysRevA.82.043803,PhysRevLett.110.234101}.

Anti-$\mathcal{PT}$ symmetry is a complementary concept to $\mathcal{PT}$ symmetry, characterized by the anti-commutation of parity and time inversion operations. Unlike $\mathcal{PT}$-symmetric systems, which typically require gain media~\cite{ozdemir2019parity}, anti-$\mathcal{PT}$ symmetry can arise naturally in dissipative systems~\cite{PhysRevA.88.053810,peng2016anti}. Various experimental systems have demonstrated anti-$\mathcal{PT}$ symmetry, including thermal and cold atoms~\cite{PhysRevLett.123.193604,PhysRevA.105.043712}, magnetic materials~\cite{PhysRevLett.125.147202,PhysRevApplied.13.014053}, electrical circuits~\cite{PhysRevLett.126.215302}, diffusive systems~\cite{li2019anti}, and optical waveguides or microcavities~\cite{PhysRevLett.124.053901,PhysRevApplied.14.044050,bergman2021observation}. Anti-$\mathcal{PT}$ symmetric systems have garnered significant interest due to their distinctive effects, such as chiral mode switching~\cite{PhysRevLett.129.273601,zhang2019dynamically} and optical energy-difference-conserving dynamics~\cite{PhysRevLett.127.083601,choi2018observation}.

Optical microcavities perturbed by nanoparticles are particularly promising for modern sensing applications~\cite{zhu2010chip,armani2007label}. The evanescent coupling between nanoparticles and the microcavity boundary induces backscattering of light in both clockwise and counterclockwise propagating modes. Under specific conditions, this effect can lead to detectable changes in the frequency splitting of whispering-gallery-mode pairs, as observed in microdisks~\cite{deych2011defect,song2010perturbation}, microspheres~\cite{PhysRevA.81.053827,PhysRevLett.99.173603}, and microtoroids~\cite{PhysRevLett.103.027406,zhu2010controlled}. Recently, achieving optical anti-$\mathcal{PT}$ symmetry breaking by spinning a linear resonator with single nanoparticle has been proposed~\cite{maayani2018flying, zhang2020breaking}. Subsequent research has focused on investigating photon blockade and slow-light effects in non-Hermitian optomechanical systems with two nanoparticles~\cite{PhysRevA.107.043715,PhysRevA.107.033507}. In this paper, we extend these studies by introducing multi-particle perturbations into an anti-$\mathcal{PT}$ symmetric system. Specifically, we explore how multi-nanoparticle perturbations can control spectral transitions and photon transmission in an anti-$\mathcal{PT}$ symmetric spinning resonator. First, we demonstrate the transition of the eigenvalue spectrum from a non-Hermitian system, satisfying anti-$\mathcal{PT}$ symmetry, to a Hermitian system through perturbation modulation. Second, we investigate the dynamical photon transmission in the quasi-closed system. Our results provide a promising approach for manipulating spectral transitions and photon transmission in lossy optical systems, which holds significant potential for the development of dissipative quantum devices.

The rest of this paper is organized as follows. In Sec.~\ref{sec:model_and_setup}, We present the system model and provide a detailed setup description. In Sec.~\ref{sec:perturbation_induced_transition}, we focus on the spectral transition from a non-Hermitian system to a quasi-closed system induced by nanoparticle perturbations. In Sec.~\ref{sec:photonic_transmission}, we analyze photon dynamics in the quasi-closed system. Finally, Sec.~\ref{sec:conclusion_and_outlook} provides a summary and outlook.
\begin{figure}
  \centering
  \includegraphics[scale = 0.23]{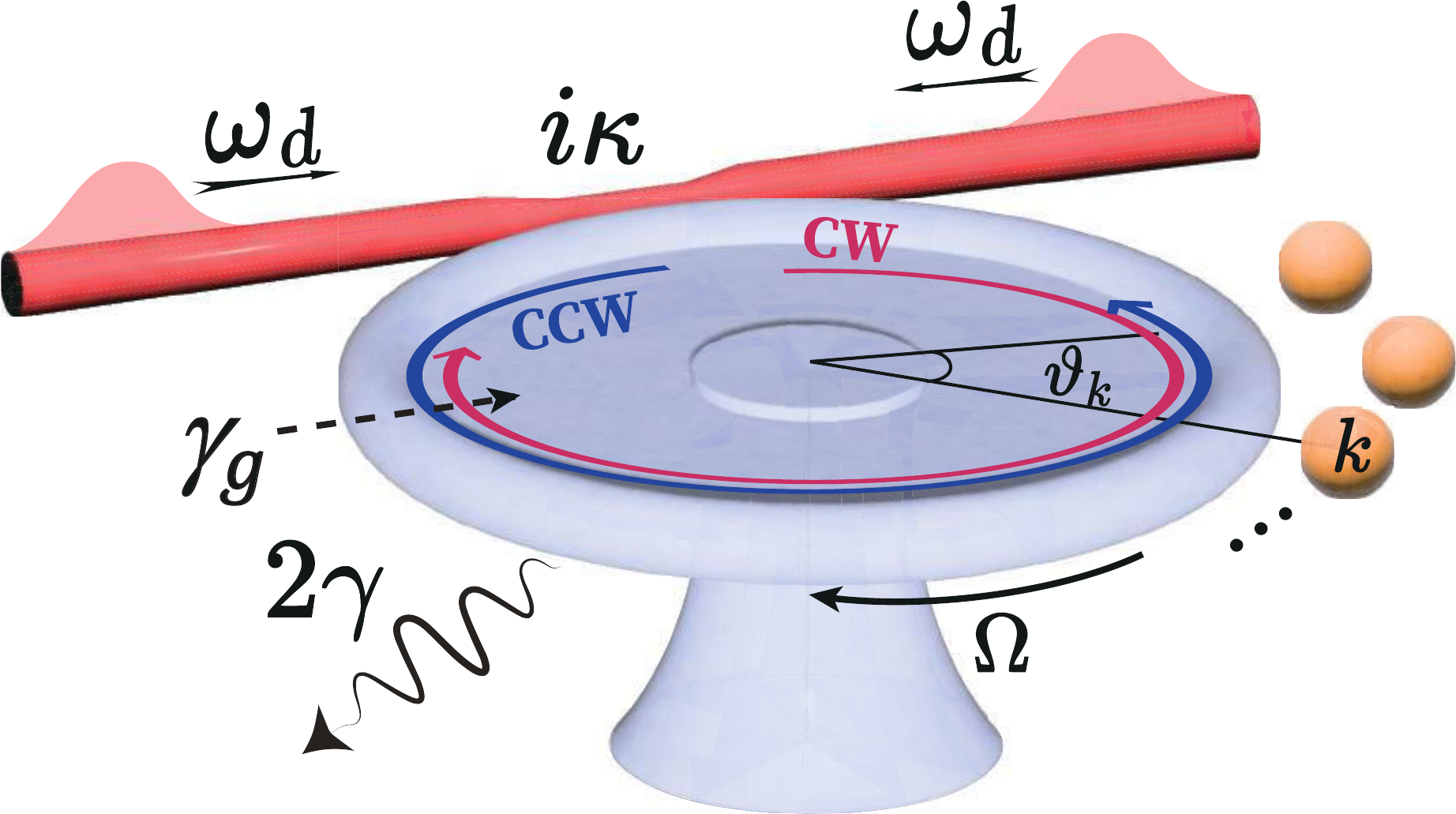}
  \caption{A linear spinning resonator with rotational angular velocity $\Omega $ is perturbed by multiple nanoparticles. Here, ${\vartheta _k}$ represents the azimuthal angle of the $k$-th nanoparticle. Two lasers with the same frequency ${\omega _d}$ drive the optical resonator from the left and right, exciting the clockwise (CW) and counterclockwise (CCW) traveling modes. The total optical loss is denoted as $\gamma  = {{\left( {{\gamma _0} + {\gamma _{c}}} \right)} \mathord{\left/
    {\vphantom {{\left( {{\gamma _0} + {\gamma _{c}}} \right)} 2}} \right.\kern-\nulldelimiterspace} 2}$, including the intrinsic loss of the optical resonator ${{\gamma _0}}$ and the loss due to the coupling of the optical resonator with fiber taper ${{\gamma _c}}$. A dissipative coupling $i\kappa$ arises from taper-scattering, which induces dissipative backscattering between the counter-circulating modes. Additionally, in the case where gain gas is introduced into the optical resonator, ${\gamma _g}$ represents the gain strength.}
  \label{Fig-modal}
\end{figure}

\section{Model and Setup}\label{sec:model_and_setup}
As illustrated in Fig.~\ref{Fig-modal}, we consider a linear optical resonator perturbed by multiple nanoparticles. The resonator is driven by two lasers with the same frequency ${\omega _d}$ from the left and right sides, exciting the clockwise (CW) and counterclockwise (CCW) traveling modes. Here, we use the biorthogonal basis ${( {\hat a_ + ^\dag ,\hat a_ - ^\dag })^T}$, where ${\hat a_ + ^\dag }$ and ${\hat a_ - ^\dag }$ represent the creation operators for the CW and CCW modes. The effective Hamiltonian in the rotating frame at frequency ${\omega _d}$ describes the linear optical resonator in the absence of rotation and nanoparticle perturbation, with $\hbar  = 1$ here and throughout, can be expressed as
\begin{align}
  {\hat H_0} = \left({
    \begin{array}{*{20}{c}}
      {{\Delta } - i\gamma } & {i\kappa }             \\
      {i\kappa }             & {{\Delta } - i\gamma }
    \end{array}
  }\right).\label{II-1}
\end{align}
The detuning $\Delta  = {\omega _0} - {\omega _d}$ represents the detuning between the optical resonator frequency ${\omega _0}$ and the driving laser frequency ${\omega _d}$. The total optical loss, $\gamma  = {{\left( {{\gamma _0} + {\gamma _{c}}} \right)} \mathord{\left/
  {\vphantom {{\left( {{\gamma _0} + {\gamma _{c}}} \right)} 2}} \right.\kern-\nulldelimiterspace} 2}$, consists of two components: the intrinsic loss of the resonator, ${\gamma _0} = {{{\omega _0}} \mathord{\left/
  {\vphantom {{{\omega _0}} Q}} \right.
  \kern-\nulldelimiterspace} Q}$, where $Q$ is the quality factor, and the loss due to coupling between the resonator and the fiber taper, ${{\gamma _c}}$. Additionally, dissipative coupling $i\kappa $ arises from taper-scattering-induced backscattering between the counter-circulating modes~\cite{lai2019observation}.

Considering the influence of multiple nanoparticles acting as Rayleigh scatterers, which either fall into or pass through the evanescent field of the resonator, these nanoparticles are fabricated by wet etching tapered fiber tips, prepared through heating and stretching standard optical fibers~\cite{jing2018nanoparticle}. The nanopositioner controls the position of each nanoparticle, allowing for adjustments in both their relative positions and effective sizes~\cite{zhu2010controlled}. Within the two-mode approximation~\cite{PhysRevA.84.063828, PhysRevLett.112.203901}, the modified effective Hamiltonian of the perturbed system, expressed in the traveling-wave basis (CW and CCW), is represented by a $2 \times 2$ non-Hermitian matrix~\cite{PhysRevA.84.063828}:
\begin{align}
  {{\hat H}_\delta } = \left(
  {
    \begin{array}{*{20}{c}}
      {\Delta  - i\gamma  + \varphi } & {i\kappa  + {\upsilon _1}}      \\
      {i\kappa  + {\upsilon _2}}      & {\Delta  - i\gamma  + \varphi }
    \end{array}
  }
  \right).\label{II-2}
\end{align}
Here, the real part of the diagonal element corresponds to the characteristic frequency of the system, while the imaginary part represents the decay rate of the resonant traveling waves. The complex off-diagonal elements ${{\upsilon _1}}$ and ${{\upsilon _2}}$ represent the backscattering coefficients describing the scattering from the CCW (CW) to the CW (CCW) wave. The backscattering between CW and CCW traveling waves is asymmetric, i.e., $\left| {{\upsilon _1}} \right| \ne \left| {{\upsilon _2}} \right|$. In the specific case of two scatters~\cite{PhysRevA.107.043715}, while ignoring frequency shifts for negative-parity modes, we have $\varphi  = \sum\nolimits_k {{\xi _k}} $, ${\upsilon _1} = \sum\nolimits_k {{\xi _k}} \exp \left( {2im{\vartheta _k}} \right)$, and ${\upsilon _2} = \sum\nolimits_k {{\xi _k}} \exp \left( { - 2im{\vartheta _k}} \right)$, where ${\xi _k} = {\varpi _k} - i{\lambda _k}$ characterizes the complex perturbation induced by the $k$-th nanoparticle, with a frequency shift ${\varpi _k}$ and linewidth broadening ${\lambda_k}$~\cite{PhysRevA.84.063828,PhysRevLett.112.203901,PhysRevApplied.10.014006}. The azimuthal mode number is denoted by $m$, and ${{\vartheta _k}}$ represents the azimuthal angle of the $k$-th nanoparticle~\cite{PhysRevA.84.063828}. By using nanopositioners to control the distance between the nanoparticle and the resonator, we can adjust the magnitudes of ${\varpi _k}$ and ${\lambda_k}$. Additionally, tuning the angle ${{\vartheta _k}}$ can drive the system to exceptional points (EPs), as has been experimentally demonstrated~\cite{peng2016chiral,chen2017exceptional}.

In order to validate the two-mode approximation under nanoparticle perturbations, both the optical spinning resonator and nanoparticles shall remain relatively stationary. In this scenario, the nanoparticles rotate synchronously with the optical resonator, both sharing the same angular velocity $\Omega $. When the system undergoes perturbation and rotates with $\Omega $, the Sagnac-Fizeau effect induces a frequency shift in the resonator, and thereby causes the frequency $\Omega_0$ to change as ${\omega _0} \to {\omega _0} \pm {\Delta _{\rm{sag}}}$, as described in ~\cite{malykin2000sagnac,PhysRevA.84.063828}
\begin{align}
  \Delta_{\text{sag}}=\frac{nR\Omega\omega_{0}}{c}\left(1-\frac{1}{n^2}-\frac{\lambda}{n}\frac{\mathrm{d}n}{\mathrm{d}\lambda}\right),\label{II-3}
\end{align}
where $n$ and $R$ correspond to the refractive index and radius of the optical resonator, respectively, and ${\omega _0} = {c \mathord{\left/{\vphantom {c \lambda }} \right.\kern-\nulldelimiterspace} \lambda }$ is the eigenfrequency of a non-spinning resonator, with $c$ and $\lambda $ denoting the speed and wavelength of light. The dispersion term $dn/d\lambda$ characterizes the relativistic origin of the Sagnac effect, which is typically negligible in conventional materials~\cite{maayani2018flying}. We define ${\Delta _{\rm{sag}}}$ and $- {\Delta _{\rm{sag}}}$ to represent the Sagnac effects in the CW and CCW traveling modes, respectively. In general, to compensate for resonator losses, a gain medium is introduced into the system \cite{changQuantumNonlinearOptics2014,aellenUnderstandingOpticalGain2022,smiraniConventionalLinearLorentzian2022}, which provides a gain of $i \gamma_g$ in both CW and CCW modes. Taking into account the rotation of the optical resonator, the inclusion of the gain medium, and the perturbation from multiple nanoparticles, the Hamiltonian of the system is rewritten as~\cite{zhang2020breaking}
\begin{align}
  {{\hat H} } = \left(
  {
    \begin{array}{*{20}{c}}
      {{\Delta _ + } - i\gamma^{'} + \varphi } & {i\kappa  + {\upsilon _1}}                \\
      {i\kappa  + {\upsilon _2}}               & {{\Delta _ - } - i\gamma^{'}  + \varphi }
    \end{array}
  } \right).\label{II-4}
\end{align}
Here, $\Delta_{\pm} =  \Delta \pm \Delta_{\text{sag}}$ are the detunings of the CW and CCW modes, including the Sagnac effect, and $\gamma^{'} = \gamma - \gamma_g$ denotes the actual loss of the system after incorporating the gain in both CW and CCW modes. Details of the derivations of Hamiltonian~(\ref{II-4}) are provided in Appendix~\ref{appendix-1}. Next, we investigate the spectral transitions of the Hamiltonian~(\ref{II-4}) by controlling the nanoparticles. The eigenvalue structure of the system will provide insights into its response to nanoparticle-induced perturbations and reveal the underlying physical mechanisms

\begin{figure}
  \centering
  \includegraphics[width=\linewidth]{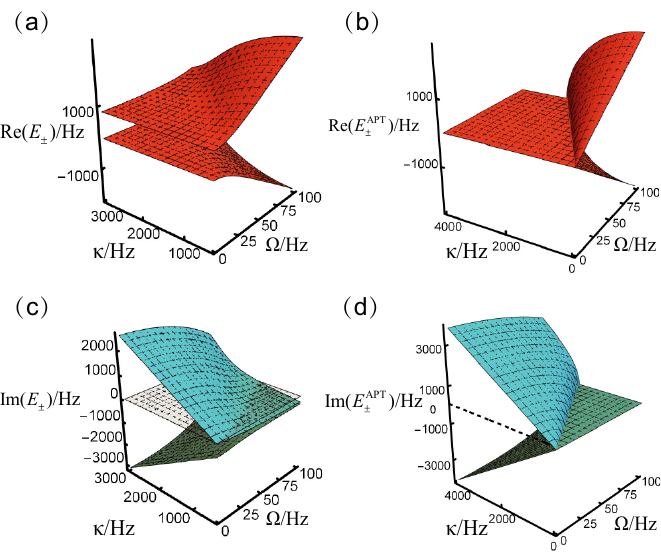}
  \caption{Phase transition behavior of a spinning optical resonator under nanoparticle perturbations. The real (red) and imaginary (cyan) parts of eigenfrequencies versus $\Omega$ and $\kappa$. $\left( \rm a \right)$ and $\left( \rm c \right)$ depict a normal perturbation scenario, with experimental parameters $\xi_1 = 1.5 \gamma -0.355 i \gamma$, $\xi_2 = 1.4\gamma - 0.645 i \gamma$, $\vartheta  = {\pi  \mathord{\left/
    {\vphantom {\pi  4}} \right.
    \kern-\nulldelimiterspace} 4}$, and $m = 4$ as in Ref.~\cite{PhysRevA.84.063828}. $\left( \rm b \right)$ and $\left( \rm d \right)$ illustrate the unperturbed case where the system satisfies Anti-$\mathcal{PT}$ symmetry. The remaining numerical simulation parameters for $\left( \rm a \right)$-$\left( \rm d \right)$ are set as $Q = {10^{12}}$, $\lambda  = 1550 \rm nm$, $n = 1.44$, $R = 50\mu \rm m$, $\gamma  = {{\left( {{\gamma _0} + {\gamma _c}} \right)} \mathord{\left/
    {\vphantom {{\left( {{\gamma _0} + {\gamma _c}} \right)} 2}} \right.
    \kern-\nulldelimiterspace} 2}$, ${\gamma _g} = 0$, ${\omega _0} = 1.94 \times {10^{14}}$Hz, $\Delta  = 0$, and ${\gamma _0} = 2{\gamma _c} = 194$Hz, following Refs.~\cite{PhysRevA.84.063828,armani2003ultra,PhysRevLett.116.133902,peng2014loss}.}
  \label{Fig-phase}
\end{figure}

\section{Perturbation-induced spectral transition}\label{sec:perturbation_induced_transition}
Without loss of generality, we consider the perturbation caused by two nanoparticles as an example, observing the spectral transition of the system from a general non-Hermitian system to one that satisfies anti-$\mathcal{PT}$ symmetry, and eventually to a quasi-closed system.

For the case of two nanoparticles, setting the azimuthal angles to ${\vartheta _1} = 0$ and ${\vartheta _2} =  - \vartheta $  modifies the Hamiltonian (\ref{II-4}) to become:
\begin{align}
  \hat H' = \left( {\begin{array}{*{20}{c}}
                        {{\Delta _ + } - i\gamma^{'}  + {\varepsilon _0}} & {i\kappa  + {\varepsilon _1}}                     \\
                        {i\kappa  + {\varepsilon _2}}                     & {{\Delta _ - } - i\gamma^{'}  + {\varepsilon _0}}
                      \end{array}} \right). \label{III-1}
\end{align}
Here, we define ${\varepsilon _0} = {\xi _1} + {\xi _2}$, ${\varepsilon _1} = {\xi _1} + {\xi _2}\exp \left( {2im\vartheta } \right)$, and ${\varepsilon _2} = {\xi _1} + {\xi _2}\exp \left( { - 2im\vartheta } \right)$. By solving the secular equation det$( {\hat H' - E\mathbb{I}} ) = 0$, where $\mathbb{I}$ is the $2 \times 2$ identity matrix, we obtain the eigenvalues and the corresponding eigenstates of the general non-Hermitian system (\ref{III-1}), yielding
\begin{align}
  {E_ \pm } =                         & \Delta  \pm i\Delta ' + {\varepsilon _0} - i\gamma^{'}, \label{III-2}                                                                  \\
  \left| {{E_ \pm }} \right\rangle  = & \frac{1}{{{C_ \pm }}}{\left( {\frac{{{\Delta _{{\rm{sag}}}} \pm \Delta '}}{{i\kappa  + {\varepsilon _2}}},1} \right)^T}, \label{III-3}
\end{align}
where $\Delta ' = {\left[ {\Delta _{{\rm{sag}}}^2 + \left( {i\kappa  + {\varepsilon _1}} \right)\left( {i\kappa  + {\varepsilon _2}} \right)} \right]^{{1 \mathord{\left/
  {\vphantom {1 2}} \right.
  \kern-\nulldelimiterspace} 2}}}$ is the normalized frequency, including the Sagnac-Fizeau effect and the influence of perturbations from two nanoparticles, and the normalized constants ${{C_ \pm }}$ are given by
\begin{align}
  {C_ \pm } = \sqrt {1 + {{\left| {{{\left( {{\Delta _{{\rm{sag}}}} \pm \Delta ' } \right)} \mathord{\left/
  {\vphantom {{\left( {{\Delta _{{\rm{sag}}}} \pm \Delta } \right)} {\left( {i\kappa  + {\varepsilon _2}} \right)}}} \right.
  \kern-\nulldelimiterspace} {\left( {i\kappa  + {\varepsilon _2}} \right)}}} \right|}^2}}.\label{III-4}
\end{align}

The eigenvalues (\ref{III-2}) demonstrate the spectral structure exhibited by the non-Hermitian Hamiltonian (\ref{III-1}) in the general case. In Figs.~\ref{Fig-phase}$\left({\rm a} \right)$ and \ref{Fig-phase}$\left({\rm c} \right)$, we present the real and imaginary components of ${E_ \pm }$, using experimental parameters from Ref.~\cite{PhysRevA.84.063828}. Figures \ref{Fig-phase}$\left({\rm a} \right)$ and \ref{Fig-phase}$\left({\rm c} \right)$ show that when the angular frequency $\Omega $ increases or the dissipation rate $\kappa $ decreases, the level splitting indicated by ${\mathop{\rm Re}\nolimits} \left( {{E_ \pm }} \right)$ intensifies, while the dissipative trend represented by ${\mathop{\rm Im}\nolimits} \left( {{E_ \pm }} \right)$ diminishes. Notably, under high-speed rotation and low dissipation, the coupling between counter-circulating modes causes the eigenfrequencies to transition from complex to real values. Next, we explore three specific cases associated with the Hamiltonian (\ref{III-1}), where two nanoparticles are manipulated to observe intriguing changes in the spectral structure of the system.

Firstly, we place two nanoparticles at positions far from the optical resonator.~When the condition $\max \left\{ {{\xi _1},{\xi _2}} \right\} \ll \min \left\{ {\gamma' ,\kappa } \right\}$ is satisfied, the perturbation from the nanoparticles can be neglected, and the Hamiltonian~(\ref{III-1}) reduces to the intrinsic form without nanoparticle-induced perturbations. When the driving laser frequency ${\omega _d}$ resonates with the inherent frequency of the system ${\omega _0}$, i.e., $\Delta  = 0$, the system is simplified to:
\begin{align}
  {\hat H_{\rm{APT}}} = \left( {\begin{array}{*{20}{c}}
                                    {{\Delta _{\rm sag}} - i\gamma^{'} } & {i\kappa }                              \\
                                    {i\kappa }                           & { - {\Delta _{\rm sag}} - i\gamma^{'} }
                                  \end{array}} \right),\label{III-5}
\end{align}
which satisfies the anticommutation relation with the $\mathcal{PT}$ operator, i.e., $\{ {{\hat H_{\rm{APT}}},\mathcal{PT}}\} = 0$, indicating that anti-$\mathcal{PT}$ symmetry can naturally exist in a linear spinning resonator driven by a laser, without requiring gain~\cite{PhysRevLett.124.053901}, nonlinearity~\cite{PhysRevLett.123.193604}, or complex spatial structures~\cite{zhang2019dynamically,peng2016anti}. What is fundamentally different from $\mathcal{PT}$ symmetry is that anti-$\mathcal{PT}$ is completely independent of the spatially separated gain-loss blanced structure. The eigenfrequencies of the anti-$\mathcal{PT}$-symmetric system are
\begin{align}
  E_ \pm ^{\rm{APT}} =  - i\gamma^{'}  \pm \sqrt {\left( {{\Delta _{\rm{sag}}} + \kappa } \right)\left( {{\Delta _{\rm{sag}}} - \kappa } \right)}.\label{III-6}
\end{align}
Here, a phase transition occurs as $\Omega $ varies. Figures \ref{Fig-phase}$\left( \rm b \right)$ and \ref{Fig-phase}$\left( \rm d \right)$ demonstrate that when ${\Delta _{\rm sag}} < \kappa $, the eigenmodes preserve anti-$\mathcal{PT}$ symmetry, maintaining identical resonance frequencies while exhibiting different linewidths. In the anti-$\mathcal{PT}$-symmetric regime, the eigenfrequencies and eigenstates are expressed as $E_ \pm ^{ \rm{APT}} =  - i( {\gamma^{'}  \mp \tilde \Delta } )$ and $| {E_ \pm ^{\rm APT}}\rangle  = {{{{( { - i{\Delta _{{\rm{sag}}}} \pm \tilde \Delta ,\kappa })}^T}} \mathord{/
  {\vphantom {{{{\left( { - i{\Delta _{{\rm{sag}}}} \pm \tilde \Delta ,\kappa } \right)}^T}} {\sqrt 2 }}} \kern-\nulldelimiterspace} {\sqrt 2 }}\kappa $, respectively, where $\tilde \Delta  = {( {{\kappa ^2} - \Delta _{{\rm{sag}}}^2} )^{{1 \mathord{\left/
  {\vphantom {1 2}} \right.
  \kern-\nulldelimiterspace} 2}}}$. In this regime, the system experiences significant dissipation because ${E_ \pm ^{\rm APT}}$ is purely imaginary. As $\Omega $ increases, an exceptional point (EP) occurs at ${\Delta _{\rm sag}} = \kappa $, where the eigenstates coalesce. At the EP, the eigenfrequency and eigenstate are ${E_{\rm{EP}}} =  - i\gamma^{'} $ and $\left| {{E_{\rm {EP}}}} \right\rangle  = {{{{\left( { - i,1} \right)}^T}} \mathord{/{\vphantom {{{{\left( { - i,1} \right)}^T}} {\sqrt 2 }}} \kern-\nulldelimiterspace} {\sqrt 2 }}$. When ${\Delta _{\rm sag}} > \kappa $, the system transitions into a phase where anti-$\mathcal{PT}$ symmetry broken, leading to bifurcating eigenmodes. The eigenfrequencies in this phase form anti-conjugate pairs, and satisfies $E_ + ^{{\rm{B - APT}}} =  - {\left( {E_ - ^{{\rm{B - APT}}}} \right)^*}$, with $E_ + ^{{\rm{B - APT}}} =  - i\gamma ' + \bar \Delta $. The corresponding eigenstates are given by
\begin{align}
  \left| {E_ \pm ^{{\rm{B - APT}}}} \right\rangle = {{{{\left( { - i{\Delta _{{\rm{sag}}}} \pm i\bar \Delta ,\kappa } \right)}^T}} \mathord{/
  {\vphantom {{{{\left( { - i{\Delta _{{\rm{sag}}}} \mp \tilde \Delta,\kappa } \right)}^T}} {C_ \pm ^ \bullet }}} \kern-\nulldelimiterspace} {C_ \pm ^ \bullet }},
\end{align}
where $\bar \Delta  = \sqrt{{\Delta _{\rm sag}^2 - {\kappa ^2}}}$ and $C_ \pm ^ \bullet  = \sqrt{{\kappa ^2} + {{\left( {{\Delta _{{\rm{sag}}}} \mp \Delta } \right)}^2}}$.

\begin{figure}
  \centering
  \includegraphics[width=\linewidth]{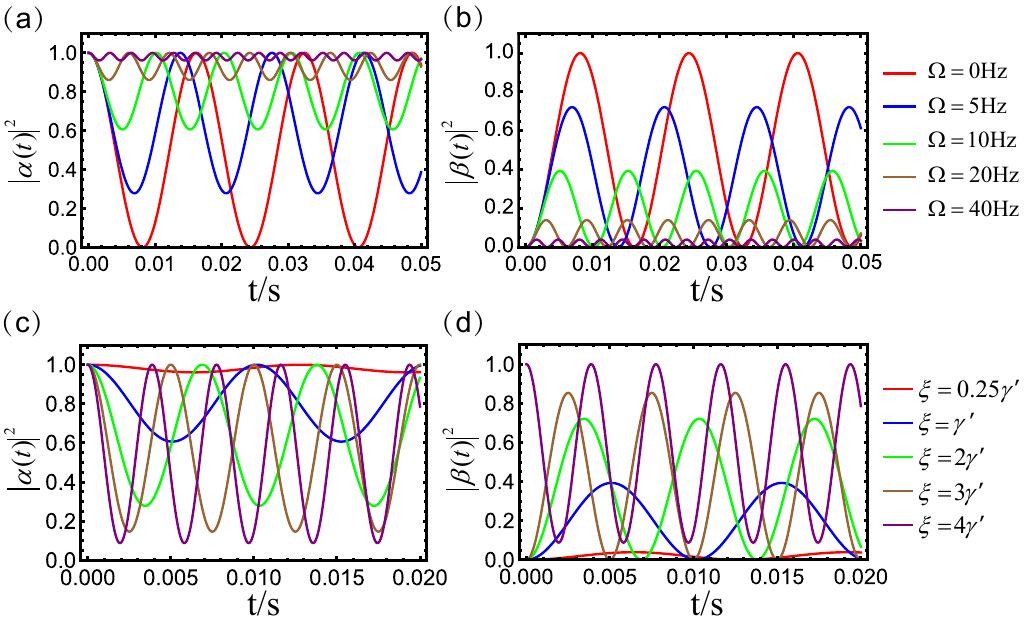}
  \caption{The dynamic photon distribution in a quasi-closed system. $\left( \rm a \right)$ and $\left( \rm b \right)$ plot the probability amplitudes of the CW and the CCW modes over time at varying angular velocities $\Omega $, with parameters $\xi  = \gamma^{'}  = 145$Hz and $\Delta  = 500$Hz. $\left( \rm c \right)$ and $\left( \rm d \right)$ show the probability amplitudes of the CW and the CCW modes over time for different perturbation strength $\xi $, where $\Omega  = 10$Hz and $\Delta  = 500$Hz. All other parameters are the same as in Fig.~\ref{Fig-phase}.}
  \label{Fig-dynamical}
\end{figure}

Secondly, while keeping one nanoparticle stationary at a distant location, we move another nanoparticle closer to the linear spinning resonator. The anti-$\mathcal{PT}$ symmetry described by Hamiltonian (\ref{III-5}) breaks immediately under the perturbation caused by a single nanoparticle. Theoretically, the sensitive responses of the anti-$\mathcal{PT}$ symmetric system to perturbations can be revealed by measuring variations in the transmission spectrum~\cite{jing2018nanoparticle,shang2023coupling}. Meanwhile, sensitivity defined by frequency splitting in experiments~\cite{hodaei2017enhanced,chen2017exceptional} can be evaluated by monitoring the separation of spectral lines in the transmission spectrum. The sensitivity of anti-$\mathcal{PT}$ symmetric systems (\ref{III-5}) to nanoparticle-induced perturbations has laid the foundation for the development of anti-$\mathcal{PT}$ sensors~\cite{zhang2020breaking}.

Thirdly, we position two nanoparticles close to the linear spinning resonator to introduce a dual-perturbation. Remarkably, intriguing effects arise when delicately manipulating two nanoparticles while simultaneously satisfying the following conditions:
\begin{align}
  {\rm{Im}}\left( {{\xi _1}} \right) = \frac{{\gamma^{'}  - \kappa }}{2}, {\rm{Im}}\left( {{\xi _2}} \right) = \frac{{\gamma^{'}  + \kappa }}{2}, \vartheta  = \frac{{2\ell  + 1}}{{2m}}\pi, \label{III-7}
\end{align}
where $\ell \in Z$. When the conditions of Eq.~(\ref{III-7}) are met, the non-Hermitian Hamiltonian (\ref{III-1}) transforms into a Hermitian Hamiltonian ${H_\mathbb{R}}$, marking a transition from an open system to a quasi-closed system, while energy is conserved. In the quasi-closed system, under stable dynamic flow equilibrium, energy is conserved. As a result, the Hamiltonian (\ref{III-1}) converts into a real form give by ${\rm{Re}}( {\hat H'})$. The eigenfrequencies and eigenstates of this quasi-closed system are represented by $E_ \pm ^ \mathbb{R} = {\omega _d} \pm \Lambda  + \xi$ and $\left| {E_ \pm ^ \mathbb{R}} \right\rangle  = {{\left( {{\Delta _{\rm sag}} \pm \Lambda ,\xi } \right)} \mathord{\left/
  {\vphantom {{\left( {{\Delta _{\rm sag}} \pm \Lambda ,\xi } \right)} {{\mathbb{C}_ \pm }\xi }}} \right.
  \kern-\nulldelimiterspace} {{ \mathbb{C}_ \pm }\xi }}$, where $\Lambda  = {\left( {\Delta _{\rm sag}^2 + \xi}  \right)^{{1 \mathord{\left/
  {\vphantom {1 2}} \right.
  \kern-\nulldelimiterspace} 2}}}$, ${\mathop{\rm Re}\nolimits} \left( {{\xi _1} + {\xi _2}} \right) = \xi $, and the normalized constants are given by
\begin{align}
  {\mathbb{C}_ \pm } = \sqrt {1 + {{\left| {{{\left( {{\Delta _{{\rm{sag}}}} \pm \Lambda } \right)} \mathord{\left/
  {\vphantom {{\left( {{\Delta _{{\rm{sag}}}} \pm \Lambda } \right)} \xi}} \right.
  \kern-\nulldelimiterspace} \xi}}\right|}^2}}.\label{III-8}
\end{align}

\section{Photonic transmission in a quasi-closed system}\label{sec:photonic_transmission}
In contrast to the energy conservation associated with an entire real spectrum under $\mathcal{PT}$-symmetry~\cite{xue2017PT}, anti-$\mathcal{PT}$ symmetry corresponds to a purely imaginary spectrum, signifying a strongly dissipative or gain-dominated process. In general, the non-Hermitian systems governed by Hamiltonians, such as Eq.~(\ref{III-1}) and Eq.~(\ref{III-5}), do not conserve energy, rendering the dynamical transfer of photon distribution inconsequential. Therefore, we focus on analyzing the dynamic transfer of photon distribution in a quasi-closed system. We restrict the quasi-closed system to the subspace spanned by the basis $\left\{ {\left| {1;0} \right\rangle ,\left| {0;1} \right\rangle } \right\}$, where $| a; b \rangle = | a \rangle \otimes |  b \rangle$, and $| n \rangle$ ($n = a,b$) represents the Fock states of the CW and CCW modes, respectively. The wave function at any time $t$ is expressed as $| \psi(t) \rangle = \alpha (t) | 1 0 \rangle + \beta (t) | 0 1 \rangle$. Under the semi-classical approximation~\cite{PhysRevA.78.053806}, quantum noise which contributes to photonic correlations \cite{PhysRevApplied.21.044048,scully1997quantum,gardiner2004quantum,PhysRevA.108.053703}, can be neglected, which allows us to focus on the dynamical behavior of the quasi-closed system by solving the Sch\"{o}dinger equation $i{\partial _t}\left| {\psi \left( t \right)} \right\rangle  = {\hat H_\mathbb{R}}\left| {\psi \left( t \right)} \right\rangle $. We assume that initially, the photon is excited in the CW mode, with $\alpha \left( 0 \right) = 1$ and $\beta \left( 0 \right) = 0$. By solving the steady-state amplitudes ${\alpha _s}\left( t \right)$ and ${\beta _s}\left( t \right)$, the probabilities of finding the photon in the CW and CCW modes can be determined
\begin{align}
  \alpha_s \left( t \right) = & \left[ \cos \left( \Lambda t \right) - \frac{\Delta_{\rm{sag}} \sin \left( \Lambda t \right)}{\Lambda} \right] \Gamma, \label{III-9} \\
  \beta_s \left( t \right) =  & - \frac{\Gamma \sin \left( \Lambda t \right) \xi}{\Lambda}, \label{III-10}
\end{align}
where $\Gamma  = \exp \left\{ { - it\left( {{\omega _d} + \xi } \right)} \right\}$. Details of the derivations of Eqs.~(\ref{III-9}) and (\ref{III-10}) are provided in Appendix~\ref{appendix-2}.

In our setup, the angular velocity of the optical resonator $\Omega$ and the nanoparticle perturbation strength $\xi  = \gamma^{'} $ are two crucial factors influencing the dynamical behavior of system. Here, we plot the probabilities $|\alpha_s\left(t\right)|^2$ and $|\beta_s\left(t\right)|^2$ as functions of time for different values of $\Omega$ and $\xi  = \gamma^{'} $. On the one hand, Figs.~\ref{Fig-dynamical}$\left( \rm a \right)$ and \ref{Fig-dynamical}$\left( \rm b \right)$ numerically demonstrate that, under the condition of a fixed perturbation $\xi  = \gamma^{'} $, the dynamical exhibit regular periodic oscillations. If the photon is initially excited in the CW mode, i.e., ${\alpha _s}\left( 0 \right) = 1$, ${\left| {{\alpha _s}\left( t \right)} \right|^2}$ first decays to zero, followed by a revival. When $\Omega  = 0$, the photon distribution between CW and CCW modes exhibits Rabi oscillations, corresponding to a swap behavior that is relevant to the charge-discharge process in quantum batteries \cite{Bechlerbatteries2018,Kamin_2020,lu2024topologicalquantumbatteries}. As $\Omega $ increases, the oscillations of ${\left| {{\alpha _s}\left( t \right)} \right|^2}$ and ${\left| {{\beta _s}\left( t \right)} \right|^2}$ become more pronounced, with shorter periods and a gradual reduction in the maximum exchange amplitude. At a higher rotational frequency of $\Omega  = 40$Hz, the photon exchange between CW and CCW modes becomes nearly isolated. This frequency-dependent control of photon dynamics can be harnessed to design non-reciprocal quantum devices, such as isolators~\cite{PhysRevLett.121.203602,shang2019nonreciprocity,PhysRevResearch.2.033517} and single-photon resources~\cite{PhysRevA.98.023856,aharonovich2016solid,sun2019optical}. On the other hand, Figs.~\ref{Fig-dynamical}$\left( \rm c \right)$ and~\ref{Fig-dynamical}$\left( \rm d \right)$ demonstrate that, when the angular velocity is fixed at $\Omega  = 10$Hz, the photon distribution becomes highly sensitive to the intensity of nanoparticle perturbations, highlighting its potential for sensor applications~\cite{zhang2020breaking,hokmabadiNonHermitianRingLaser2019,PhysRevLett.128.173602}. The above analysis demonstrates that fine-tuning both $\Omega$ and $\xi = \gamma^{'}$ allows for precise control over the photon distribution between the CW and CCW modes, offering potential for the design and implementation of versatile quantum devices.

\section{Conclusion and outlook}\label{sec:conclusion_and_outlook}
In this work, we have explored the impact of nanoparticle-induced perturbations on the energy spectrum of an anti-$\mathcal{PT}$ symmetric system. In the absence of perturbations, the system initially exhibited anti-$\mathcal{PT}$ symmetry. However, the introduction of a single nanoparticle quickly broke this symmetry, demonstrating the potential of this sensitivity for the development of anti-$\mathcal{PT}$ sensors. Notably, when multiple nanoparticles and a gain medium were introduced, the non-Hermitian system transitioned into a Hermitian one, forming an energy-conserving quasi-closed system. By investigating the dynamics of such systems, we showed that precise control over the angular velocity of the optical resonator and the strength of nanoparticle perturbations enabled the design of various quantum devices, including energy storage and non-reciprocal devices. Looking ahead, exploring the dynamical behavior near exceptional points in time-dependent anti-$\mathcal{PT}$ symmetric quantum systems represents a promising direction for future research~\cite{Wang:24}. Additionally, investigating the symmetry and higher-order exceptional points in anti-$\mathcal{PT}$ topological materials could provide further exciting avenues for development~\cite{PhysRevLett.127.186601}.

\acknowledgements
C. Shang acknowledges financial support from the China Scholarship Council, the Japanese Government (Monbukagakusho-MEXT) Scholarship under Grant No. 211501, the RIKEN Junior Research Associate Program, and the Hakubi projects of RIKEN.

\appendix
\section{Details of the derivation of Eq.~(\ref{II-4})}\label{appendix-1}

The optical resonator model comprises rotatable optical cavities and fibers with mutual coupling. By introducing two laser driving fields with a frequency $\omega_d$ on either side of the fiber, clockwise (CW) and counter-clockwise (CCW) traveling modes can be excited within the optical cavity from the left and right, respectively. When the laser driving frequency deviates from the intrinsic frequency of the cavity, the driving laser induces detuning for both propagation modes ($a_{cw}$ and $a_{ccw}$) inside the optical cavity. The detuning magnitude is given by $\Delta = \omega_0 - \omega_d$. The intrinsic frequency of the optical cavity is given by $\omega_0 = c/\lambda$, where $c$ is the speed of light and $\lambda$ is the wavelength of the light. The optical cavity has intrinsic losses, and the loss rate is $\gamma_0 = \omega_0 / Q$, where $Q$ is the quality factor representing the resistance of the optical cavity to loss. Additionally, coupling losses occur between the optical cavity and the fiber, characterized by $\gamma_c$. The propagation modes in both directions, $a_{cw}$ and $a_{ccw}$, experience the same loss rate $\gamma = (\gamma_0 + \gamma_c)/2$. The scattering-induced backscattering between the CW and CCW modes results in dissipative coupling in the fiber, characterized by a magnitude of $i\kappa$ \cite{lai2019observation,
  hokmabadiNonHermitianRingLaser2019}. In this setting, the Hamiltonian of the system is written as
\begin{align}
  \mathcal{\hat H}_0' & = \left(\omega_0 - i \gamma\right) \hat a_+^{\dag}\hat a_+  + i \kappa_0 \hat a_-^{\dag} \hat a_+ + i \kappa_0 \hat a_+^{\dag} \hat a_- \nonumber \\
                      & + \left(\omega_0 - i \gamma \right) \hat a_-^{\dag} \hat a_-,
\end{align}
where \(\hat a_+\) (\(\hat a_-\)) and \(\hat a_+^{\dag}\) (\(\hat a_-^{\dag}\)) represent the annihilation and creation operators for the CW (CCW) mode photons, respectively. By introducing \(\mathcal{\hat H}_0' = \varphi^\dag \hat{H}_0' \varphi\), where \(\varphi = (\hat a_+, \hat a_-)^T\), the effective Hamiltonian of the system in the ${\sigma _z}$ representation is given by $\hat{H}_0' = (\omega_0 - i\gamma) \mathbb{I} + i\kappa \hat\sigma_x$, where $\mathbb{I}$ is the $2 \times 2$ identity matrix, and ${\hat\sigma _x}$ is the Pauli matrix. Here, the CW and CCW modes are represented by the basis vectors \((1, 0)\) and \((0,1)\). As the microcavity rotates in the system, we adopt a rotating reference frame with frequency ${\omega _d}$ and define the transformation matrix $U = e^{i\omega_d t} \mathbb{I}$. After applying the unitary transformation, the effective Hamiltonian \(\hat{H}_0\) of the system in the ${\sigma _z}$ representation becomes \cite{zhang2020breaking} $\hat{H}_0 = (\Delta - i \gamma) \mathbb{I} + i \kappa \hat\sigma_x$, which is identical to Eq. (\ref{II-1}) in the main text.

Since the size of the nanoparticles (typically around 10 nm) is much smaller than the wavelength of the driving laser in the optical cavity (approximately 1550 nm in the model), they meet the Rayleigh scattering condition, where $r_j \ll \lambda$. Consequently, we treat the nanoparticles as Rayleigh scatterers, either traversing the evanescent field of the resonator or falling into it. The nanoparticles couple only to the fixed modes of the optical microdisk~\cite{PhysRevA.84.063828}, so we analyze their effect on the optical modes within the standing wave basis. In the Rayleigh scattering limit, perturbations caused by nanoparticles at $x = 0$ do not affect the odd-parity modes in the standing wave basis. Thus, the perturbation due to a single nanoparticle can be expressed in the standing wave basis as
$\mathcal{\hat H}_p = 2\xi \hat a^\dagger_{\text{even}} \hat a_{\text{even}}$,
where $\xi = \varpi - i \lambda_d$ characterizes the strength of the nanoparticle perturbation. Here, $\varpi$ denotes the frequency shift introduced by the nanoparticle, reflecting its coupling strength with the system, while $\lambda_d$ represents the spectral linewidth of the nanoparticle, accounting for its dissipation effects~\cite{PhysRevA.84.063828,PhysRevLett.112.203901,PhysRevApplied.10.014006}. In the ${\sigma _z}$ representation, the effective Hamiltonian of the nanoparticle in the standing wave basis, for even parity (1,0) and odd parity (0,1), is written as $\hat{H}_p = \xi (\mathbb{I} + \hat \sigma_z)$, where ${\hat \sigma _z}$ is the Pauli matrix, \( \mathcal{\hat H}_p = \phi^\dag \hat{\hat H}_p \phi \), and \( \phi = \left(\hat a_{\text{even}}, \hat a_{\text{odd}} \right)^T \).

To account for the effects of nanoparticles positioned at different azimuthal angles, it is necessary to transform the nanoparticle perturbations from the standing wave basis to the traveling wave basis. The traveling wave basis states, characterized by specific propagation directions and phase angles, can be seen as superpositions of the standing wave basis states. Consequently, the creation and annihilation operators for the two propagation directions in the traveling wave basis can be expressed as linear combinations of the odd and even mode operators in the standing wave basis,
\begin{align}
   & {a_ + } = \frac{1}{{\sqrt 2 }}\left( {{a_{{\rm{even}}}}{e^{im\varphi }} + i{a_{{\rm{odd}}}}{e^{im\varphi }}} \right),       \\
   & {a_ - } = \frac{1}{{\sqrt 2 }}\left( {{a_{{\rm{even}}}}{e^{ - im\varphi }} - i{a_{{\rm{odd}}}}{e^{ - im\varphi }}} \right).
\end{align}
Here, the nanoparticle perturbation at the azimuthal angle \( \varphi = \theta \) can be transformed into the traveling wave basis, consisting of CW \(\left(1, 0\right)\) and CCW \(\left(0, 1\right)\), through the transformation matrix
\begin{equation}
  M = \frac{1}{\sqrt{2}} \left(\begin{array}{cc}
      e^{i m\varphi}  & i e^{i m\varphi}   \\
      e^{-i m\varphi} & -i e^{-i m\varphi}
    \end{array}\right).
\end{equation}
The transformed effective Hamiltonian for the nanoparticle perturbations in the traveling wave basis is given by
\begin{equation}
  \hat{H}_p' = M \hat{H}_p M^\dag = \xi \left(\begin{array}{cc}
      1              & e^{2im\theta} \\
      e^{-2im\theta} & 1
    \end{array}\right).
\end{equation}
By neglecting the extremely weak interactions between nanoparticles, the effective Hamiltonian of multiple nanoparticle perturbations is given by \cite{PhysRevA.84.063828}
\begin{equation}\label{Eq:multi-pert}
  \hat{H}_\delta' = \sum\limits_k \xi_k\left(\begin{array}{cc}
      1                   & e^{2 i m \theta_k} \\
      e^{-2 i m \theta_k} & 1
    \end{array}\right),
\end{equation}
where \( \xi_k \) represents the effect of the \( k \)-th nanoparticle on the system, \( \theta_k \) denotes the azimuthal angle of the \( k \)-th nanoparticle, and \( m \) is the azimuthal mode number of the nanoparticles \cite{PhysRevA.84.063828}.

Under the two-mode approximation \cite{PhysRevLett.112.203901,jing2018nanoparticle}, where there is only one mode in each of the CW and CCW directions, the total effective Hamiltonian of the system, perturbed from the traveling-wave basis (CW and CCW), can be expressed as follows \cite{PhysRevA.84.063828}:

\begin{equation}\label{Eq:h-not-rotating}
  \hat{H}_0 + \hat{H}_\delta' = \left(\begin{array}{cc}
      \Delta_d - i \gamma + \sum\limits_k \xi_k        & i \kappa  + \sum\limits_k \xi_k e^{2im\theta_k} \\
      i \kappa  + \sum\limits_k \xi_k e^{-2im\theta_k} & \Delta_d - i \gamma + \sum\limits_k \xi_k
    \end{array}\right).
\end{equation}

To ensure the validity of the two-mode approximation under nanoparticle perturbations Eq.~(\ref{Eq:h-not-rotating}), we keep the nanoparticle and the micro-ring cavity relatively stationary. A nanopositioner is used to ensure that the nanoparticle and the micro-ring cavity rotate together with an angular velocity $\Omega$ on the base. When the entire micro-ring cavity rotates clockwise with angular velocity $\Omega$, this rotation induces a Sagnac-Fizeau effect within the resonator \cite{malykin2000sagnac}:
\begin{equation}\label{Eq:Delta_sag}
  \Delta_{\text{sag}}=\frac{nR\Omega\omega_0}{c}\left(1-\frac{1}{n^2}-\frac{\lambda}{n}\frac{\mathrm{d}n}{\mathrm{d}\lambda}\right),
\end{equation}
where $n$ and $R$ represent the refractive index and the microdisk radius, respectively. Here, $\omega_0 \rightarrow \omega_0 \pm \Delta_{\text{sag}}$ represents the eigenfrequency of a non-rotating resonator, where $c$ and $\lambda $ denote the speed of light and its wavelength, respectively. The inherent frequencies \( \omega_0 \) in the two directions of the micro-ring cavity undergo opposite shifts, i.e., \( \omega_0 \rightarrow \omega_0 \pm \Delta_{\text{sag}} \). For classical materials, the dispersion term \( \frac{dn}{d\lambda} \), which accounts for relativistic effects in the Sagnac frequency shift, is typically very small (less than 1\%) \cite{maayani2018flying} and can be neglected in the calculations presented here. The optical detuning for the propagation modes in the CW and CCW directions is given by ${\Delta _{{\rm{cw}}}} = {\Delta _ + } = {\Delta _d} + {\Delta _{{\rm{sag}}}}$ and ${\Delta _{{\rm{ccw}}}} = {\Delta _ - } = {\Delta _d} - {\Delta _{{\rm{sag}}}}$, respectively. Here, $\Delta_{\text{sag}}$ represents the Sagnac frequency shift, which occurs in opposite directions due to the rotation of resonator, while $\Delta = \omega_0 - \omega_d$ denotes the optical drive detuning, which is the same in both directions and induced by the laser drive.

By introducing a gain medium into the optical system, a gain of \( i\gamma_g \) with equal magnitude in both traveling modes can be produced \cite{changQuantumNonlinearOptics2014,aellenUnderstandingOpticalGain2022,smiraniConventionalLinearLorentzian2022}. Finally, the total effective Hamiltonian of the system in the traveling-wave basis reads:
\begin{equation}\label{Eq:h-eff}
  \hat{H} = \left(\begin{array}{cc}
      \Delta_+ - i \gamma^{'} + \sum\limits_k \xi_k    & i \kappa  + \sum\limits_k \xi_k e^{2im\theta_k} \\
      i \kappa  + \sum\limits_k \xi_k e^{-2im\theta_k} & \Delta_- - i \gamma^{'} + \sum\limits_k \xi_k
    \end{array}\right),
\end{equation}
where $\gamma^{'} = \gamma - \gamma_g$. We ultimately reproduce the same equations as  Eq.~(\ref{II-4}) presented in the main text.

\section{Details of the derivation of Eqs.~(\ref{III-9}) and (\ref{III-10})}\label{appendix-2}

Under the semi-classical approximation~\cite{PhysRevA.78.053806}, the time evolution of the photonic distribution in the system can be derived. By substituting the wavefunction $|\psi (t)\rangle  = \alpha (t)|10\rangle  + \beta (t)|01\rangle $ into the Sch\"{o}dinger equation $i{\partial _t}\left| {\psi \left( t \right)} \right\rangle  = \hat{H}_{\mathbb{R}}\left| {\psi \left( t \right)} \right\rangle$, the steady-state amplitude ${\alpha _s}\left( t \right)$ and ${\beta _s}\left( t \right)$ can be obtained explicitly by solving the equations
\begin{align}
   & i\frac{{d{\alpha _s}\left( t \right)}}{{dt}} = \left( {{\Delta _ + } + \xi } \right){\alpha _s}\left( t \right) + {\beta _s}\left( t \right)\xi = 0, \nonumber            \\
   & i\frac{{d{\beta _s}\left( t \right)}}{{dt}} = {\alpha _s}\left( t \right)\xi  + \left( {{\Delta _ - } + \xi } \right){\beta _s}\left( t \right) = 0. \label{steady-state}
\end{align}
To solve Eqations~(\ref{steady-state}), we define the transformation matrix \( R_{\mathbb{R}} = \left( | E^{\mathbb{R}}_+ \rangle, | E^{\mathbb{R}}_- \rangle \right) \) and the diagonal matrix \( D_{\mathbb{R}} \) = diag$\left[ {E^{\mathbb{R}}_+,E^{\mathbb{R}}_-} \right]$, where \( | E^{\mathbb{R}}_+ \rangle \) and \( | E^{\mathbb{R}}_- \rangle \) are the eigenstates of the quasi-closed system, with corresponding eigenvalues \( E^{\mathbb{R}}_+ \) and \( E^{\mathbb{R}}_- \), respectively. The transformation matrix \( R_{\mathbb{R}} \) composed of eigenstates, possesses the following properties: $H_{\mathbb{R}} = R_{\mathbb{R}} D_{\mathbb{R}} R_{\mathbb{R}}^{-1}$ and $D_{\mathbb{R}} = R_{\mathbb{R}}^{-1} H_{\mathbb{R}} R_{\mathbb{R}}$,
where \( R_{\mathbb{R}} \) can convert the diagonal matrix \( D_{\mathbb{R}} \) containing the Hamiltonian eigenvalues into the Hamiltonian \( H_{\mathbb{R}} \). Conversely, it also diagonalizes \( H_{\mathbb{R}} \) into the matrix \( D_{\mathbb{R}} \).

Thus, we can utilize the transformation matrix \( R_{\mathbb{R}} \) to reformulate the Sch\"{o}dinger equation $i{\partial _t}{\left[ {{\alpha _s},{\beta _s}} \right]^T} = {H_{\mathbb{R}}}{\left[ {{\alpha _s},{\beta _s}} \right]^T}$ into a more tractable form. By defining $R_{\mathbb{R}}^{-1} \left(\alpha_s, \beta_s \right)^T = \left(x, y \right)^T$, we obtain
\begin{equation}\label{Eq:dynamic-diagonal}
  i\frac{d}{dt} \left(x, y \right)^T = D \left(x, y \right)^T.
\end{equation}
The solutions to Eq.~(\ref{Eq:dynamic-diagonal}) are $ x  = c_1 \exp\left(i E^{\mathbb{R}}_+ t\right)$ and $ y = c_2 \exp\left(i E^{\mathbb{R}}_- t\right)$, where \(c_1\) and \(c_2\) are constants determined by the initial conditions. By multiplying both sides of $R_{\mathbb{R}}^{-1} \left(\alpha_s, \beta_s \right)^T = \left(x, y \right)^T$ by \(R_{\mathbb{R}}\), we obtain \(R_{\mathbb{R}}R_{\mathbb{R}}^{-1} \left(\alpha_s, \beta_s \right)^T = R_{\mathbb{R}} \left(x, y \right)^T\). Since \(R_{\mathbb{R}} R_{\mathbb{R}}^{-1} = 1\), the inverse transformation yields \(\left(\alpha_s, \beta_s \right)^T = R_{\mathbb{R}} \left(x, y \right)^T\). For the initial conditions $\alpha (0) = 1$ and $\beta (0) = 0$, we have $R_{\mathbb{R}} \left( c_1,c_2 \right) = \left(1, 0 \right)$, which gives the values of ${c_1} = {a \mathord{\left/
  {\vphantom {a {2\Lambda }}} \right.
  \kern-\nulldelimiterspace} {2\Lambda }}$ and ${c_2} =  - {a \mathord{\left/
  {\vphantom {a {2\Lambda }}} \right.
  \kern-\nulldelimiterspace} {2\Lambda }}$, where $\Lambda  = {\left( {\Delta _{\rm sag}^2 + \xi}  \right)^{{1 \mathord{\left/
  {\vphantom {1 2}} \right.
  \kern-\nulldelimiterspace} 2}}}$. Thus far, we have reproduced Eqs.~(\ref{III-9}) and (\ref{III-10}) as presented in the main text.

\bibliography{reference}

\end{document}